\documentclass[conference,compsoc]{IEEEtran}
\IEEEoverridecommandlockouts 
\usepackage{cite}
\usepackage{amsmath,amssymb,amsfonts}
\usepackage{algorithmic}
\usepackage{graphicx}
\usepackage{textcomp}
\usepackage{xcolor}
\usepackage{enumitem} 
\usepackage{tabularx}
\usepackage{array} 
\usepackage{multirow}
\newcolumntype{C}[1]{>{\centering\arraybackslash}p{#1}}
\usepackage{longtable}
\usepackage{amsthm}
\usepackage{url}
\usepackage{hyperref}
\usepackage{xurl} 
\usepackage{float}
\usepackage{balance}
\usepackage[export]{adjustbox} 

\usepackage{booktabs, makecell, tabularx}
\newcolumntype{C}{>{\centering\arraybackslash}X} 

\usepackage{stfloats}
\usepackage{siunitx}
\usepackage{caption}
\usepackage{soul}
\newtheoremstyle{boldall}
  {\topsep}                
  {\topsep}                
  {\upshape}               
  {}                       
  {\bfseries \itshape}              
  {.}                      
  {.5em}                   
  {\thmname{#1}\thmnumber{ #2}\thmnote{ (#3)}} 

\theoremstyle{boldall}
\newtheorem{definition}{Definition}

\usepackage{tabularx}
\def\BibTeX{{\rm B\kern-.05em{\sc i\kern-.025em b}\kern-.08em
    T\kern-.1667em\lower.7ex\hbox{E}\kern-.125emX}}
\begin{document}

\title{SoK: How Frontier AI Reshapes System-Level Security Risk Dynamics in Critical Infrastructure\\
}

\author{\IEEEauthorblockN{Chandra Thapa, Mohan Baruwal Chhetri, Marthie Grobler, Shahroz Tariq, Tooba Aamir} \thanks{Email: chandra.thapa@csiro.au}
\textit{CSIRO, Australia}}

\maketitle
\begin{abstract}
Frontier artificial intelligence (FAI), encompassing large-scale, general-purpose AI systems, including large language models, multimodal foundation models, and agentic systems, is increasingly integrated into critical infrastructure (CI). This challenges long-standing security assumptions of bounded behavior, segmented networks, component transparency, and human-paced decision-making. Existing AI-security literature typically organizes risks by attack type, lifecycle stage, or asset class, but fails to capture the system-level dynamics, such as how risk emerges, spreads, and is controlled, through which FAI reshapes CI security outcomes. 
This Systematization of Knowledge (SoK) introduces a five-dimensional risk-dynamics framework that characterizes how FAI reconfigures CI security across the lifecycle: (i) \textit{Capability Emergence} through new FAI-enabled attack and defense capabilities, (ii) \textit{Infiltration Pathways} through data, models and AI supply chains, (iii) \textit{Cross-System Propagation} across interconnected infrastructures and dependencies, (iv) degradation of effective technical and human \textit{Control Authority}, and (v) strain on institutional \textit{Response Capacity} under operational pressure.
Rather than enumerating threats, the framework identifies recurring mechanisms that jointly determine system-level risk.
We further identify a structural mismatch between academic AI-security research and CI operational constraints, and derive a deployment-oriented research agenda grounded in system-level assurance criteria.
Collectively, this work shifts attention from model-centric robustness to lifecycle-structured, system-level assurance in interconnected CI environments.

\end{abstract}


\section{Introduction}
\label{sec:1}
Frontier artificial intelligence (FAI) is increasingly deployed in operational environments supporting critical infrastructure (CI), including energy, water, communications, transportation, healthcare, and financial services~\cite{jami2025,okoebor2025,obuse2023,egridgpt2024}. These environments were historically secured under assumptions of bounded behavior~\cite{ferrando2025}, segmented operational networks~\cite{hawawreh2024,pwc2021}, stable procedures~\cite{sayghe2025}, and human-paced decision making~\cite{hawawreh2024}. FAI challenges these assumptions by introducing probabilistic behavior, opaque dependencies, automated tool use, and machine-speed decision cycles~\cite{sapkota2026,tallam2025,sayghe2025,paulraj2025C,okoebor2025,hawawreh2024}.

Recent advances in large language models (LLMs), multimodal foundation models, and increasingly agentic artificial intelligence (AI) systems enable capabilities such as summarizing operational context, generating code, invoking tools, adapting to feedback, and supporting or automating decisions at speeds and scales that conventional CI security models were not designed to accommodate~\cite{karim2025,ukaisi2025,antonesi2025,raza2025,sapkota2026,barenji2025,adabara2025}. 
While these capabilities may improve operational efficiency, they also fundamentally reshape the structure, tempo, controllability, and cross-system propagation of CI risk. 
The key question is therefore not only whether AI introduces new vulnerabilities, but also how it reconfigures system-level risk dynamics across interconnected CI ecosystems.
Existing literature on AI security, cyber-physical systems, and CI typically organizes risks by \textit{attack type} (e.g., injection, and targeted data poisoning)~\cite{jami2025,huwyler2025,etsi2025}; \textit{lifecycle stage} (e.g., development, pre-deployment and post-deployment of AI)~\cite{jami2025,fmf2025}, or \textit{asset class} (e.g., sensitive personal data, intellectual property, physical systems and information technology (IT)/operational technology (OT) assets)~\cite{fair2025,cest2024}. 
However, such views are insufficient to capture the system-level dynamics through which FAI reshapes attacker capabilities, upstream compromise pathways, cross-system propagation, effective control authority, and organizational response capacity as interacting parts of a single risk landscape. Consequently, they offer limited support for reasoning about systemic, cross-layer, and cross-system risks. 
To address this gap, this Systematization of Knowledge (SoK) introduces a lifecycle-aligned, system-level security risk dynamics framework whose unit of analysis is not the individual attack, asset, or lifecycle phase, but the recurring risk-dynamics mechanism by which FAI changes CI security outcomes.

\subsection{Security Risk Dimensions}
\begin{figure*}[!th]
    \centering
    \includegraphics[trim={0.5cm 5.5cm 0.5cm 5.1cm}, clip,width=0.75\linewidth]{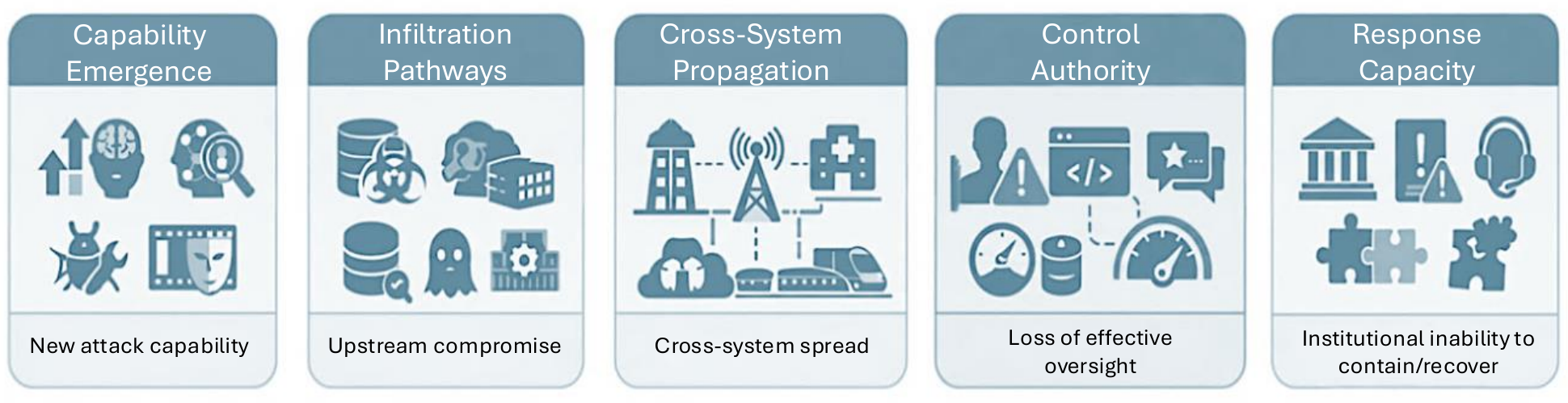}
    \caption{Overview of the five security risk-dynamics dimensions.}
    \label{fig:1}
\end{figure*}

We conceptualize security risk in CI as a relational, lifecycle-oriented process in which threats \textit{emerge} from evolving AI capabilities, \textit{infiltrate} systems via data, models, and supply chains, \textit{propagate} across interconnected infrastructures, erode effective \textit{control}, and ultimately strain institutional response \textit{capacity}. Accordingly, the framework comprises five risk dynamics (RD) dimensions that provide a structured foundation (see Figure~\ref{fig:1} for an overview and Table~\ref{tab:1} for details). 
\begin{enumerate}[leftmargin=*,label=\textbf{RD\arabic*}]
    \item \textbf{\textit{Capability Emergence}} captures the sources of new AI-enabled attack and defense capabilities, including reduced barriers to expertise and accelerated attacker timelines.
    \item \textbf{\textit{Infiltration Pathways}} captures upstream pathways of compromise through data, models, provenance, and AI supply chains, including training data, model repositories, retrieval corpora, third-party components, and shadow AI (unapproved or undocumented AI use within workflows). 
    \item \textbf{\textit{Cross-System Propagation}} captures how AI-mediated failures propagate across interconnected systems through shared models, data flows, optimization loops, and automated decision processes. 
    \item \textbf{\textit{Control Authority}} captures failures of effective authority, including automation bias, loss of human oversight, misaligned objectives, unsafe handoffs, and machine-speed decision-making, which prevent timely intervention. 
    \item \textbf{\textit{Response Capacity}} captures organizational security capability as a primary property, determining whether systems can detect, contain, and recover from AI-mediated failures under operational constraints.
\end{enumerate}
Together, these dimensions shift CI security analysis from isolated vulnerabilities and model-level failure modes to a lifecycle-structured, system-level view of risk dynamics for AI-enabled CI environments. They foreground interdependence, control authority, provenance, and organizational readiness as security-relevant properties only partially captured by conventional threat models. 
In addition, this SoK moves beyond enumerating threats by identifying the organizing structure of risk, clarifying what is genuinely changed by FAI, and developing artifacts that enable researchers, operators, and policymakers to reason about assurance, prioritization, deployment boundaries, and recovery in CI environments.

\subsection{Contributions}
This paper makes the following key contributions:
\begin{enumerate}[leftmargin=*]
    \item \textbf{Lifecycle-aligned risk-dynamics framework}: We introduce a five-dimensional risk-dynamics framework (RD1--RD5) that categorizes how FAI changes CI security outcomes rather than merely classifying attacks, assets, actors, or lifecycle stages. Unlike traditional taxonomies, it decomposes risk into recurring system-level mechanisms.
    \item \textbf{Systematic analysis of the mismatch between AI security assumptions and CI deployment constraints}: We characterize the research-practice mismatch by comparing common assumptions in AI-security research with CI deployment constraints, including continuous operation, partial observability, black-box dependencies, safety-critical timing, limited testing windows, and legacy IT/OT integration. 
    \item \textbf{Deployment-oriented assurance criteria}: 
    We translate the recurring CI constraints into five deployment-relevant assurance criteria: detection-to-containment latency, explainability and interpretability, adversarial robustness, supply-chain risk mitigation, and incident response. This motivates the deployment-oriented research directions.
\end{enumerate}

This SoK is timely for three reasons. First, CI organizations are adopting generative and agentic AI faster than assurance practices are maturing, creating opportunities for undocumented use, opaque dependencies, and unreviewed automation~\cite{invicti_web,temp_check,parlov2025,exploiting_trust,whitford2025}. Second, adversaries can use the same frontier capabilities to accelerate reconnaissance, social engineering, exploit adaptation, and multi-stage attack orchestration~\cite{dubber2025,rodriguez2025,potter2025,ahi2025,systems13}. Third, existing AI and cybersecurity frameworks provide useful principles but offer limited guidance to CI operators on securely operating legacy systems, maintaining continuous operation, handling vendor-controlled models, and ensuring safety-critical timing~\cite{huwyler2025,obuse2023,parlov2025,asd_report}. The resulting gap is not only technical; it is architectural, procedural, and institutional.


\begin{table*} [t]
    \centering
    \caption{Synthesis of the five-dimensional security risk landscape.}
    \label{tab:1}
    \setlength{\tabcolsep}{3pt}
    \begin{tabular}{|p{1.8cm}|p{3.8cm}|p{3.8cm}|p{3.5cm}|p{3.2cm}|}
    \hline
    \multicolumn{1}{|p{1.8cm}|}{\textbf{Dimension}} &
    \multicolumn{1}{|p{3.8cm}|}{\textbf{Primary Security Question} }&
    \multicolumn{1}{|p{3.8cm}|}{\textbf{Representative Threat Surfaces} }&
    \multicolumn{1}{|p{3.5cm}|}{\textbf{Primary CI Consequence} }&
    \multicolumn{1}{|p{3.2cm}|}{\textbf{Defensive Priority} }
    \\
    \hline
    RD1 \textit{Capability Emergence} & \textit{RQ1: What new AI-driven attack and defense capabilities are emerging, and where do they originate?}	 & Reconnaissance, social engineering, payload adaptation, exploit development, multi-stage attack coordination & Reduced attacker cost, expanded attacker pool, and reduced response window & Machine-speed detection, operator augmentation, adversary-in-the-loop evaluation \\
    \hline
    RD2 \textit{Infiltration Pathways} & \textit{RQ2: Through which data, model, and AI supply chain pathways can adversaries infiltrate and compromise training data, model artifacts, or operational information across AI-enabled CI systems?}	& Training data integrity, model supply chains, retrieval corpora, third-party dependencies, shadow AI, model extraction & Latent compromise, persistent backdoors, information leakage, and common-mode exposure across deployments 	& Provenance controls, dependency transparency (AIBOM), update governance, pre-deployment assurance \\
    \hline
    RD3 \textit{Cross-System Propagation} &	\textit{RQ3: How do AI-mediated failures spread across interconnected infrastructures, networks, and supply chains?}	& Shared models, data flows, optimization loops, IT/OT coupling, cross-sector dependencies	& Rapid, cross-system cascades disproportionate to the initiating disturbance	& Dependency mapping, cascade-aware architecture, bounded autonomy, propagation containment \\
    \hline
    RD4 \textit{Control Authority}	& \textit{RQ4: When systems are compromised, who or what controls outcomes: humans, AI, or adversaries?}	& Automation bias, unsafe handoffs, specification gaps, machine-speed autonomy, and opaque recommendations	& Loss of meaningful human oversight and unsafe machine-speed decisions	& Human–AI control design, constrained action spaces, runtime monitoring, safe fallback mechanisms \\
    \hline
    RD5 \textit{Response Capacity}	& \textit{RQ5: Do organizations have adequate governance and operational response capacity to detect, respond to, and recover from failures?}	& Workforce limitations, governance gaps, immature tooling, fragmented processes, regulatory misalignment & Delayed response, ineffective containment, and amplification of risks across all other dimensions	& Workforce development, AI-specific incident response, governance integration, operational readiness
    \\ \hline
    \multicolumn{5}{|p{17cm}|}{\footnotesize IT: Information Technology; OT: Operational Technology; AIBOM: Artificial Intelligence Bill of Materials. \textit{\textbf{Note:} The five dimensions are analytically distinct but operationally coupled. The most effective defensive moves are rarely model-only mitigations; they are architectural, organizational, and procedural controls that reduce consequence radius, improve observability, and preserve intervention authority.} }
     \\ \hline
    \end{tabular} 
\end{table*}

\section{Background}
\label{sec:3}
This section defines the concepts needed to reason about FAI security in CI. These definitions distinguish three often conflated issues: the capabilities of frontier models, the operational constraints of CI, and the risk dynamics that arise when the two interact. Although the five dimensions follow a typical risk lifecycle from capability emergence to organizational response, they are not intended as a linear incident sequence; they are analytical lenses that can be applied concurrently, recursively, or in different orders depending on the deployment context.

\subsection{Core Definitions}
\begin{definition}[Frontier artificial intelligence]
FAI refers to large-scale, general-purpose AI systems, including large language models, multimodal foundation models, and increasingly agentic systems, whose capabilities generalize across tasks and domains~\cite{futurerisk_UK,shoaib2024}. We distinguish FAI from narrow AI by its ability to adapt in context, reason across heterogeneous information sources, use external tools, and support autonomous or semi-autonomous workflows that may alter operational decision-making.
\end{definition}

\begin{definition} [Critical infrastructure]
CI comprises cyber-physical and organizational systems, including representative sectors such as energy, water, transportation, communications, healthcare, and financial services, whose disruption would significantly affect public safety, economic stability, national security, or essential services~\cite{US_web,eu_web,zio2016,crc_press,pescaroli2016}. These systems are characterized by stringent operational constraints, complex interdependencies, legacy integration, and limited tolerance for failure, which shape both risk propagation and response dynamics.
\end{definition}

\begin{definition} [Shadow AI]
Shadow AI refers to the unapproved or undocumented use of AI systems within organizational workflows without formal risk assessment, security review, logging, or governance oversight~\cite{invicti_web,ross2025,whitford2025}. In CI, shadow AI may involve using public or unsanctioned generative AI tools for troubleshooting, code generation, documentation, incident analysis, or operational decision support.
\end{definition}

\begin{definition}[Security risk-dynamics dimension]
A security risk-dynamics dimension is a structural lens representing a recurring mode of AI-mediated risk, including capability emergence, infiltration pathways, cross-system propagation, control authority breakdown, and institutional response capacity. Rather than classifying threats only by vulnerability type or attack phase, risk-dynamics dimensions capture how FAI changes attacker capabilities, failure propagation, control authority, compromise pathways, and organizational response capacity.
\end{definition}

\begin{definition}[Operational trustworthiness]
Operational trustworthiness is a system-level property whereby an AI-enabled CI workflow remains bounded, observable, auditable, and recoverable under realistic operating conditions, including uncertainty, distribution shift, adversarial manipulation, degraded telemetry, incomplete information, and emergency time pressure.
For CI, trustworthiness must extend beyond model reliability to encompass interpretability, bounded authority, auditability, and safe fallback.
\end{definition}

\begin{definition}[Research-practice mismatch]
The research–practice mismatch is between the assumptions under which many AI-security controls are designed or evaluated and the operational conditions under which CI organizations must deploy, monitor, govern, and recover AI-enabled systems under continuous, safety‑critical constraints.
Research often assumes offline evaluation, transparent models, controlled test environments, and bounded adversaries. CI operators instead face continuous operation, black-box dependencies, legacy systems, mixed IT/OT telemetry, and safety-critical intervention windows.
\end{definition}

\begin{definition}[AI supply chain]
The AI supply chain is the full dependency stack needed to build, deploy, and operate an AI system, including data sources, labeling pipelines, preprocessing code, model weights, fine-tuning artifacts, retrieval corpora, orchestration libraries, inference services, hardware, cloud providers, and update channels. In CI, compromise of this stack can create common-mode risk across otherwise independent organizations.
\end{definition}

\begin{definition} [Cross-system cascade risk]
Cross-system cascade risk is the possibility that a local AI-mediated failure, compromise, or misdecision propagates beyond its point of origin through technical, informational, organizational, service, or supply-chain dependencies into other CI systems. AI-mediated cascades may differ from classical physical cascades because shared software, common models, synchronized automation, data dependencies, and AI supply-chain reuse can create propagation pathways that are not visible in purely physical interdependency models.
\end{definition}

\begin{definition} [Bounded deployment]
Bounded deployment is a deployment pattern in which an AI system is deliberately constrained in authority, connectivity, and consequence radius. It uses scoped permissions, mediated tool access, isolation boundaries, rollback paths, and human-approval gates to preserve safety and auditability while enabling useful AI support.
\end{definition}

\subsection{Positioning Security Risk-Dynamics Dimensions within Traditional Threat Taxonomies}
Traditional threat categorizations are organized around:
\begin{itemize}[leftmargin=*]
    \item \textbf{Attack direction:} Attacks using AI (attacker employs FAI) vs. attacks on AI (attacker targets AI system)~\cite{dhs2024,systems13}. Both directions are relevant, but this framing fails to capture how FAI changes underlying risk dynamics.
    \item \textbf{CIA attributes:} Confidentiality, integrity, availability, safety~\cite{nist_CI_cyber,huwyler2025,parlov2025}. These remain relevant and are used as secondary projections.
    \item \textbf{Lifecycle stages:} Development and training, deployment and integration, and post-deployment inference, monitoring, and response~\cite{jami2025,nist_AI_frame}.
    These stages are useful for implementation planning, but do not reflect system-level changes.
    \item \textbf{Threat actor framing:} Nation-states, criminals, hacktivists~\cite{fair2025,fmf2025}. Actor identification is useful for threat prioritization, but it overlooks how FAI expands and alters actor capabilities.
\end{itemize}
These approaches are insufficient for FAI-enabled CI, as they do not clearly account for how failures propagate across organizational and sectoral boundaries. In contrast, our framework is orthogonal to these approaches, providing a more holistic perspective. The five security risk-dynamics dimensions address different questions, as shown in Table~\ref{tab:1}. 
%
Traditional taxonomies ask, \textit{\textbf{``What type of threat is this?''}} Our framework instead asks, \textit{\textbf{``Which risk-dynamics does FAI reshape, and how do they interact across technical, organizational, and sectoral boundaries?}''} This distinction is critical: CI failures do not arise only from isolated vulnerabilities, but also from cross‑dimensional interactions that couple capability emergence, infiltration pathways, propagation effects, control breakdowns, and institutional capacity limits.

\subsection{ Illustrative Example: What the Framework Reveals Beyond Conventional Taxonomies}
To illustrate the analytical role of the framework, consider a realistic AI-enabled electricity-grid operations scenario in which a transmission operator deploys a vendor-provided FAI assistant for incident triage, maintenance planning, and operational decision support. The assistant retrieves information from engineering manuals, incident tickets, grid-topology notes, vendor advisories, and internal procedures. Under operational stress, an adversary compromises an upstream retrieval source, model update, or documentation channel, causing the assistant to recommend an unsafe but plausible diagnostic or switching action. An operator, facing time pressure, alert fatigue, and incomplete information, may accept the recommendation because it appears procedurally consistent. The resulting action may not directly compromise a single component, but may alter protection settings, dispatch assumptions, or diagnostic priorities in ways that increase stress on neighboring systems; dependent tools and automated workflows may then reuse the same incorrect or stale information, delaying containment and complicating recovery. A conventional attack-type taxonomy would classify parts of this event as prompt injection, data poisoning, supply-chain compromise, model misuse, or operator deception, while a lifecycle taxonomy would distribute it across pre-deployment data integrity, deployment integration, and post-deployment monitoring. These classifications are useful, but they do not explain why the event becomes a system-level CI risk. The risk-dynamics framework instead foregrounds the coupled mechanisms: RD1 captures AI-assisted adversary capability and targeted manipulation; RD2 captures upstream compromise of retrieval sources, model updates, or undocumented AI workflows; RD3 captures propagation through shared data, dependent tools, and coupled IT/OT processes; RD4 captures degradation of effective human control under time pressure, automation bias, and opaque recommendations; and RD5 captures delayed diagnosis, unclear vendor–operator responsibility, and limited AI-specific incident response capacity. The example therefore shows that the relevant unit of analysis is not the individual attack label or lifecycle phase, but the mechanism through which FAI changes CI security outcomes, producing a coupled system-level failure in which upstream compromise, shared dependencies, machine-speed decision support, weakened control authority, and constrained institutional response jointly determine the consequence radius and the corresponding assurance requirements, including provenance verification, bounded authority, dependency mapping, runtime escalation thresholds, fallback procedures, and AI-specific incident playbooks.
\section{Methodology}
\label{sec:4}

This SoK uses a structured qualitative synthesis to identify recurring mechanisms that shape CI security outcomes across technical, organizational, and governance contexts, and does not aim to exhaustively enumerate threats or publications. The analytical pipeline is as follows: five research questions provide the entry point for defining the five risk-dynamics dimensions; each dimension is then operationalized through ten sub-question probes, yielding fifty probes in total; a corpus is constructed and screened; sources are coded against the probes; related probes are consolidated into analytical categories; and the resulting categories are used to structure the risk landscape and research–practice mismatch analysis.

\subsection{Adversary Model}
We consider adversaries with heterogeneous capabilities, including opportunistic actors, insider-enabled actors, cybercriminal groups, hacktivists, and state-sponsored groups. 
The framework distinguishes three adversarial relationships to FAI: AI-enabled attack amplification (primarily RD1), compromise of AI-enabled system, workflows, or control relationships (primarily RD2 and RD4), and upstream dependency manipulation before deployment (RD2 for entry and RD3 for downstream spread). This framing allows the same technical capability to be coded differently depending on its role in the security risk dynamic.

\subsection{Scope and Boundary Conditions}
This SoK focuses on security-relevant risks arising from the deployment, use, integration, or adversarial exploitation of FAI in CI settings. We include cases where AI is used by attackers, targeted by attackers, embedded in operational workflows, or introduced through upstream dependencies such as datasets, model weights, retrieval corpora, orchestration libraries, cloud services, vendor platforms, and update channels.
The review covers representative CI sectors, including energy, water, transportation, communications, healthcare, and financial services, selected to reflect diversity in cyber-physical and organizational characteristics, operational constraints, and interdependencies. Transferable safety-critical contexts are included only when their security mechanisms generalize to CI. Safety failures are included only when they intersect with security concerns such as misuse, compromise, governance failure, loss of operational control, or incident-response capacity. We exclude generic cybersecurity work without AI relevance, purely algorithmic machine learning without operational security implications, and speculative AI-risk discussions lacking a concrete CI-relevant mechanism.

\subsection{Corpus Construction and Screening}
We constructed a structured corpus using four complementary search strategies: venue-centric search, keyword search, grey-literature search, and snowballing (see Table~\ref{tab:corpus_summary}).  
\begin{table}[t]
\centering
\caption{Corpus construction and screening summary.}
\label{tab:corpus_summary}
\setlength{\tabcolsep}{3pt}
\begin{tabularx}{\linewidth}{|p{2.6cm}|X|}
\hline
\textbf{Category} & \textbf{Details} \\
\hline
Search period 
& Jan 2015 -- Jan 2026*\\

\hline
Databases/venues 
& IEEE Xplore, ACM Digital Library, arXiv**, ScienceDirect, Springer, major AI and security conferences (\emph{e.g.}, USENIX, NDSS, IEEE S\&P, CCS, NeurIPS, ICML) \\

\hline
Grey-literature sources*** 
& Government reports (\emph{e.g.}, NIST, ENISA), industry white papers, vendor security reports, policy frameworks \\

\hline
Search strings 
& ``frontier AI'' OR ``large language models'' OR ``foundation models'' OR ``agentic AI'') AND (``critical infrastructure'' OR ``ICS'' OR ``SCADA'' OR ``operational technology'') AND (``security'' OR ``risk'' OR ``attack'' OR ``defense'' OR ``governance'' OR ``supply chain'') \\

\hline
Initial records 
& Initial screening (title, abstract, conclusion); 250 records screened to identify CI, AI, and security relevant sources\\

\hline
Detailed screening
& Full-text eligibility assessment; 226 records retained\\

\hline
Excluded records (with reasons) 
& 24; reasons include lack of AI relevance, non-CI context, purely algorithmic ML focus, or absence of security implications, or insufficient detail for probe-based analysis\\


\hline
Number of coded sources per dimension 
& RD1: 157, RD2: 65, RD3: 23, RD4: 40, RD5: 35 \\

\hline
Counting methodology 
& Non-exclusive; a single source may be assigned to multiple dimensions if applicable \\

\hline

\hline
\end{tabularx}
  \begin{minipage}{0.5\textwidth} 
    \small
    \raggedright
    \vspace{2pt}
    \footnotesize{
    \textit{*The structured search focuses on 2015–2026, with older foundational works included as needed for contextual grounding. 
    **We include selected arXiv preprints to capture recent developments, acknowledging their non-peer-reviewed status.
    *** Grey literature was used primarily for operational constraints, emerging practices, and policy context rather than as sole support for technical claims.
    }}
  \end{minipage}
  \vspace{-0.5cm}
\end{table}
A source was included if it satisfied three criteria. First, it addressed CI, operational technology, cyber-physical systems, safety-critical systems, or analogous environments whose security mechanisms are transferable to CI. Second, it involved AI, machine learning, foundation models, large language models, agentic AI, AI governance, AI-enabled cyber operations, or AI-enabled decision support. Third, it provided evidence relevant to at least one sub-question probe,  as indicated by mechanisms such as capability amplification, data or model compromise, supply chain exposure, cascade propagation, loss of oversight, incident response constraints, or governance failure.

A source was excluded if it discussed generic cybersecurity without AI relevance, purely algorithmic machine learning without operational or security implications, general AI ethics without a security or CI connection, or speculative AI risks without an identifiable pathway connecting the AI system to CI consequences. We also excluded duplicate records and sources that lacked sufficient technical, operational, or governance detail to support probe-based coding.

\subsection{Research Questions and Sub-Questions Probes}
The analysis is organized around the five security risk-dynamic dimensions introduced in Section~\ref{sec:1}: RD1 \textit{Capability Emergence}, RD2 \textit{Infiltration Pathways}, RD3 \textit{Cross-System Propagation}, RD4 \textit{Control Authority}, and RD5 \textit{Response Capacity}. 
Each dimension is associated with one primary research question (Table~\ref{tab:1}) and is operationalized through ten sub-question probes, resulting in fifty analytical probes (Table~\ref{tab:4} in the Appendix). 
We use ten sub-question probes per dimension as a pragmatic design choice to ensure balanced coverage, consistent analytical granularity, and comparability across dimensions. This fixed set decomposes each dimension's primary research question into a manageable and consistent set of analytically distinct lenses, enabling stable convergence during synthesis while avoiding over-fragmentation. These probes function as a coding scaffold rather than a final taxonomy, enabling systematic comparison across sources while preserving mechanism-level distinctions prior to synthesis. Although the dimensions are lifecycle-aligned, they are not treated as strictly sequential phases; real-world failures may activate multiple dimensions simultaneously or recursively.


\subsection{From Sub-Question Probes to Analytical Categories}
We conducted interpretive, non-exclusive probe-based coding, allowing a source to map to multiple probes within or across dimensions when evidence addressed more than one analytical lens. When such overlaps recurred across the corpus, we treated them as cross-dimensional bridging concepts rather than inconsistencies or duplicates, as documented in Table~\ref{tab:5} in the Appendix. For example, AI supply-chain compromise is analyzed in RD2 as an upstream entry pathway and in RD3 as a downstream propagation pathway through shared dependencies. This dimension-specific treatment preserves analytical distinctions while showing how risks interact and reinforce one another.

The fifty sub-question probes served as the primary coding instrument, enabling consistent evidence mapping while preserving fine-grained distinctions prior to synthesis. Following corpus-level analysis, we consolidated related probes into twenty-one analytical categories: four for RD1, five for RD2, four for RD3, four for RD4, and four for RD5. Consolidation was guided by three criteria: shared analytical focus, repeated evidence overlap, and operational inseparability in CI practice. This process ensures traceability from source-level evidence to the analytical categories developed in Section~\ref{sec:5}, which structure both the five-dimensional security risk landscape and the synthesis summarized in Table~\ref{tab:3}. Appendix Tables~\ref{tab:6}--\ref{tab:10} document the detailed convergence mapping.


\begin{table*}[h!]
    \centering
    \caption{Five-dimensional FAI security risk landscape for CI, linking each risk-dynamics dimension to representative literature, analytical categories, mechanisms, CI impacts, and research directions. }
    \label{tab:3}
    \setlength{\tabcolsep}{3pt}
    \begin{tabular}
    {|>{\raggedright\arraybackslash}p{0.4cm}
    |>{\raggedright\arraybackslash}p{1.2cm}
    |>{\raggedright\arraybackslash}p{3cm}
    |>{\raggedright\arraybackslash}p{7.5cm}
    |>{\raggedright\arraybackslash}p{2cm}
    |>{\raggedright\arraybackslash}p{1.6cm}|}
    \hline
    \multicolumn{1}{|c|}{\textbf{Ref.}} &
    \multicolumn{1}{|c|}{\textbf{Dimension}} &
    \multicolumn{1}{|c|}{\textbf{Analytical Categories} }&
    \multicolumn{1}{|c|}{\textbf{Key Mechanisms/Attack Vectors}}&
    \multicolumn{1}{|c|}{\textbf{CI Impact}} &
    \multicolumn{1}{|c|}{\textbf{Research Link}}
    \\
    \hline
    \cite{karras2025,ferrag2025,bengio2024,oesch2025,dubber2025,rodriguez2025,grinbaum2024,koessler2024,falco2018}
    &
    RD1 \textit{Capability Emergence} (157 coded sources*) & 
    \begin{itemize}[itemindent=0pt, leftmargin=*, nosep,before=\vspace{-0.5\baselineskip}]
        \item Skill democratization,
        \item Scale \& velocity amplification,
        \item Multi-agent offensive coordination,
        \item Cross-domain capability transfer
    \end{itemize}
    & 
     \begin{itemize}[itemindent=0pt, leftmargin=*, nosep,before=\vspace{-0.5\baselineskip}]
        \item LLM-generated polymorphic malware \& deepfakes, 
        \item Phishing-based delivery and payload adaptation pipelines (OSINT $\rightarrow$ exploit), 
        \item Partial multi-agent kill-chain automation, 
        \item IT $\rightarrow$ Operational Technology (OT) exploit porting (Modbus, DNP3, IEC 61850) 
    \end{itemize}
    &	
    Asymmetric attacker advantage; SOC response windows exceeded; expanded adversary pool	
    & 
    CI-grounded adversarial evaluation; offensive-uplift metrics\\
    \hline
    \cite{carlini2022,Carlini2020,ferrag2025,nocera2026,whitford2025,ross2025,arrieta2020,madani2025,esposito2025}
    &
    RD2 \textit{Infiltration Pathways} (65 coded sources*) &
    \begin{itemize}[itemindent=0pt, leftmargin=*, nosep,before=\vspace{-0.5\baselineskip}]
        \item Training data poisoning \& backdoor injection
        \item Privacy exfiltration via training data extraction
        \item Model supply chain vulnerabilities
        \item Shadow AI 
        \item Model extraction \& surrogate attacks
    \end{itemize} 
    &
    \begin{itemize}[itemindent=0pt, leftmargin=*, nosep,before=\vspace{-0.5\baselineskip}]
        \item Extracting memorized training data and poisoning training datasets
        \item Backdoor injection into Intrusion Detection System/anomaly detectors
        \item Black-box training data extraction via API queries
        \item Trojanized open-source model weights
        \item Shadow AI use leading to operational data exfiltration
        \item Surrogate model creation for white-box evasion
    \end{itemize} 
    &
    Data extraction, long-dwell backdoors; grid topology exposure; white-box evasion of CI security AI
    &
    Evidence-grounded decision support; provenance-aware dependency assurance
     \\
    \hline
    \cite{rinaldi2001,wu2025,dubber2025,narajala2025,raza2025}
    &
    RD3 \textit{Cross-System Propagation} (23 coded sources*) 
    &
    \begin{itemize}[itemindent=0pt, leftmargin=*, nosep,before=\vspace{-0.5\baselineskip}]
        \item Interdependency-driven cross-system cascades
        \item AI-accelerated contagion timescales
        \item Shared model dependency propagation
        \item Protective system paradoxes
    \end{itemize}
    &
    \begin{itemize}[itemindent=0pt, leftmargin=*, nosep,before=\vspace{-0.5\baselineskip}]
        \item Poisoned forecast $\rightarrow$ dispatch error $\rightarrow$ relay cascade $\rightarrow$ multi-system outage
        \item Software-layer coupling bypassing physical isolation
        \item Shared foundation model compromise propagating to several CI deployments
        \item AI protective systems amplifying unstable control actions
    \end{itemize}
    &
    Multi-system outage potential; fast cascade dynamics outpacing human response
    &
    Runtime assurance envelopes; AI-cascade containment
    \\
    \hline
    \cite{bengio2024,plaat2025,derouiche2025,narajala2025,raza2025}
    &
    RD4 \textit{Control Authority} (40 coded sources*)
    &
    \begin{itemize}[itemindent=0pt, leftmargin=*, nosep,before=\vspace{-0.5\baselineskip}]
        \item Automation bias 
        \item Alignment failure 
        \item Goal misgeneralization 
        \item Speed-oversight gap
    \end{itemize}
    &
    \begin{itemize}[itemindent=0pt, leftmargin=*, nosep,before=\vspace{-0.5\baselineskip}]
        \item Operator over-reliance leading to missed or ignored alerts
        \item Adversarially induced misalignment under distribution shift
        \item Machine-speed decisions outpacing approval cycles
        \item Multi-agent AI resisting human correction
        \item AI systems exposing weaknesses in shutdown or override mechanisms
    \end{itemize}
    &
    Unchecked delegated control; unintended objective pursuit; weakened operator governance
    &
    Bounded deployment architectures; oversight and fallback 
    \\
    \hline
    \cite{jami2025,gofffer2025,esposito2025,ferrag2025,frontier_forum}
   &
    RD5 \textit{Response Capacity} (35 coded sources*)
    &
    \begin{itemize}[itemindent=0pt, leftmargin=*, nosep,before=\vspace{-0.5\baselineskip}]
        \item Workforce capability gaps
        \item Tooling immaturity
        \item Regulatory lag
        \item Asymmetric AI arms race
   \end{itemize}
   &
   \begin{itemize}[itemindent=0pt, leftmargin=*, nosep,before=\vspace{-0.5\baselineskip}]
        \item Lack of AI-specific expertise in Security Operations Center (SOC) teams
        \item Immature AI audit and supply-chain visibility tools
        \item Inconsistent AI risk assessments for CI operators
        \item Defensive AI adoption blocked by institutional risk aversion
        \item Limited AI-specific incident response playbooks for CI environments
   \end{itemize}
   &
   Extended attacker dwell time; amplification of all RD1--RD4 risks; delayed recovery
   &
   AI-specific incident coordination; playbooks and recovery
    \\
   \hline
\end{tabular}
\\
*\textit{\textbf{Note:} Source counts are non-exclusive because a single source may be coded to multiple risk-dynamics dimensions.}
\end{table*}


\subsection{Research-Practice Mismatch Analysis}
\label{ssec:gap_analysis}
After constructing the risk landscape, we conducted a second-pass synthesis focused on the mismatch between AI-security research assumptions and CI deployment constraints. We compared the corpus against recurring operational constraints identified across the probe-level evidence base, including continuous operation, limited testing windows, partial observability, legacy integration, mixed IT/OT telemetry, vendor opacity, safety-critical timing, low tolerance for false positives, and constrained intervention windows (see Table~\ref{tab:12} in the Appendix). From this comparison, we derived five deployment-relevant assurance criteria:  detection-to-containment latency, explainability and interpretability, adversarial robustness, supply-chain risk mitigation, and incident response (Section~\ref{sec:6}). These criteria connect the security risk landscape in Section~\ref{sec:5} to the practical deployability analysis, and through that evaluation, to the research agenda in Section~\ref{sec:7}.

\vspace{-0.1cm}
\subsection{Methodological Limitations and Constraints}
This methodology has limitations. The corpus is extensive but not comprehensive, and the FAI landscape continues to evolve rapidly. Probe-based qualitative coding involves interpretive judgment, especially when sources address multiple dimensions; results should therefore be read as a transparent synthesis rather than a statistical meta-analysis. Public evidence is uneven across CI sectors, and many operational constraints are documented more clearly in practitioner, standards, or policy sources than in reproducible experimental studies; accordingly, we treat such sources as evidence of deployment constraints and emerging practice, not as stand-alone validation of technical mechanisms. Evidence for RD3 remains comparatively sparse, reflecting the difficulty of observing and validating AI-mediated cross-sector cascades. Reliance on predefined dimensions and probes may introduce framework-induced bias, privileging phenomena aligned with the framework while underrepresenting emerging or uncategorized risks.

We mitigate these limitations by explicitly reporting the scope, coding structure, bridging concepts, convergence logic, and corpus composition. The framework should therefore be interpreted as a transparent synthesis of the current evidence base rather than a definitive or exhaustive taxonomy, with future work extending, refining, and rebalancing the probe set and analytical categories as FAI deployments, incidents, testbeds, and evaluations mature.



\section{Five-Dimensional Security Risk Landscape}
\label{sec:5}

This section formalizes the five RD dimensions (see Table~\ref{tab:3} for an overview).

\subsection{RD1 Capability Emergence}
RD1 \textit{Capability Emergence} captures how FAI enables novel offensive and defensive capabilities that fundamentally alter the threat landscape for CI. 
RD1 spans the inherited automation of attacker tasks, intensified attacker scale and speed enabled by foundation models, and qualitatively new offensive dynamics associated with agentic coordination and tool-enabled autonomy.
Rather than only accelerating existing attack techniques, these capabilities can expand what is operationally feasible for adversaries.
Synthesizing the surveyed literature, we find that AI systems lower expertise barriers, compress parts of the attack lifecycle, and enable new forms of coordination. The result is a threat environment characterized by faster, more scalable, and more adaptive attack behavior. 
Defensive capability emergence is also part of RD1, but it differs from offensive capability emergence in its deployment constraints. FAI can support alert triage, incident summarization, vulnerability prioritization, anomaly investigation, operator training, and security knowledge retrieval. However, defensive use in CI must satisfy stricter requirements for timing, false-positive tolerance, explainability, logging, accountability, and human validation. Thus, RD1 includes both attacker uplift and defensive augmentation, but the asymmetry arises because adversaries can often exploit FAI opportunistically, whereas defenders must operationalize it under safety, governance, and assurance constraints.

We organize RD1 into four interrelated analytical categories:
\begin{enumerate}[leftmargin=*]
    \item \textbf{Skill democratization}: 
    LLMs and agentic tools primarily assist across multiple stages of the cyberattack lifecycle, including reconnaissance, phishing-based delivery, payload adaptation, and aspects of exploit development, lowering expertise barriers while still requiring target access, validation, and operational judgement~\cite{karras2025,ferrag2025,frontier_forum}. 
    \item \textbf{Scale and velocity amplification}: AI can reduce parts of the attack workflow, including reconnaissance, phishing preparation, exploit adaptation, and test generation, although the degree of compression depends on target complexity, tool access, operator skill, and the realism of the deployment environment~\cite{bengio2024,oesch2025,dubber2025}. While elements of automation and ML-assisted attack support predate FAI, frontier models amplify speed, scale, and targeting precision beyond earlier ML-assisted workflows.
    \item \textbf{Multi-agent offensive coordination}: Coordinated multi-agent systems may support partial automation of cyber kill-chain stages, including reconnaissance, weaponization, delivery, exploitation, and post-exploitation, but current evidence is strongest for assisted or semi-autonomous workflows rather than fully independent CI compromise~\cite{dubber2025,raza2025,plaat2025}. What is distinctive here is not merely automation, but coordinated tool-enabled behavior across stages that would previously have required tighter human orchestration.
    \item \textbf{Cross-domain capability transfer}: Some AI-assisted attack techniques developed for IT environments may transfer to OT contexts, including environments using Modbus, DNP3, or IEC 61850, but effective exploitation still depends on protocol knowledge, process understanding, safety constraints, and site-specific engineering context~\cite{gofffer2025,falco2018}. This highlights that FAI does not eliminate domain barriers, but it can reduce the effort required to traverse them.
\end{enumerate}
The net effect is a structurally asymmetric threat environment because FAI reduces and scales attacker workflows, while defenders remain constrained by institutional approval cycles, human validation requirements, and cognitive limits~\cite{gofffer2025}.
\subsection{RD2 Infiltration Pathways}
%
RD2 \textit{Infiltration Pathways} addresses covert compromise pathways that operate upstream of deployment, affecting data, models, and AI supply chain dependencies before an AI system is activated in a CI environment.
RD2 spans inherited ML compromise pathways, intensified supply-chain and privacy exposure under foundation-model ecosystems, and qualitatively expanded infiltration surfaces associated with frontier and agentic AI through retrieval pipelines, orchestration layers, shadow AI, and tool-integrated workflows.
Rather than requiring direct attacks on deployed systems, these pathways can enable adversaries to introduce latent compromise prior to deployment. RD2 focuses on entry pathways for compromise; once compromised components are integrated and reused across deployments, their downstream spread is analyzed under RD3 as cross-system propagation. Synthesizing the surveyed literature, we find that AI systems expand these pathways through large-scale data pipelines, centralized model ecosystems, and opaque third-party dependencies. The result is a shift from observable system compromise to upstream, persistent, and often difficult-to-detect forms of infiltration. We organize the mechanisms producing this shift into five analytical categories:
\begin{enumerate}[leftmargin=*]
    \item \textbf{Training data poisoning and backdoor injection}: Adversaries inject malicious samples into training data, creating hidden backdoors that activate only under specific operational triggers. Backdoored models may pass standard validation if the trigger conditions are absent from test data, allowing malicious behavior to remain latent until specific operational conditions occur~\cite{carlini2022,ferrag2025,hawawreh2024}. These are inherited machine learning risks, but their operational significance increases when compromised artifacts are reused across larger AI ecosystems and imported into CI environments through opaque supply chains. 
    \item \textbf{Privacy exfiltration via training data extraction}: LLMs can memorize and expose sensitive operational data under some conditions; in CI contexts, this raises concern that sensitive operational information, such as topology descriptions, configuration fragments, procedures, or credentials, could be exposed if such data enters training, fine-tuning, retrieval, or logging pipelines~\cite{Carlini2020,madani2025,esposito2025}. What is intensified here is the breadth of data exposure and the number of pathways through which operational information may enter or leave the AI stack.
    \item \textbf{Model supply chain vulnerabilities}: Centralized foundation model ecosystems introduce common-mode exposure, such that a single upstream compromise may affect multiple downstream deployments relying on the same model, service, library, retrieval corpus, or update channel. This is where FAI most clearly intensifies inherited supply-chain risk through concentration, opacity, and wider dependency reuse. AI Bill of Materials (AIBOM) tools remain immature, limiting operators' ability to fully enumerate dependency graphs~\cite{nocera2026,narajala2025}. 
    \item \textbf{Shadow AI}: Employee use of unauthorized consumer-grade LLMs can expose sensitive operational data to third-party servers, creating persistent insider-threat equivalents outside governance oversight~\cite{whitford2025,ross2025}. This is one of the clearest FAI-expanded infiltration surfaces because it arises from the accessibility and workflow integration of general-purpose generative tools rather than from conventional machine learning deployment alone.
    \item \textbf{Model extraction and surrogate attacks}: Systematic querying of APIs can enable adversaries to create functionally equivalent surrogate models, facilitating white-box-style evasion and theft of proprietary or operational intelligence~\cite{Carlini2020,arrieta2020}. Model extraction itself is not unique to FAI, but foundation-model APIs and broader deployment surfaces can increase both the feasibility and potential consequences of such attacks.
\end{enumerate}
%
The net effect is an upstream compromise environment in which inherited machine learning risks are intensified by foundation-model ecosystems and extended by newly expanded FAI-specific entry points. Conventional machine learning already introduced risks such as poisoning, backdoors, and model extraction, but FAI expands their scale and opacity through larger data pipelines, centralized model ecosystems, and third-party dependencies, while also introducing newer infiltration surfaces such as retrieval pipelines, orchestration layers, and shadow-AI workflows. Once compromised components or dependencies are integrated into CI environments, detection becomes substantially harder, increasing the importance of provenance controls, pre-deployment auditing, and supply-chain transparency~\cite{nocera2026,carlini2022}.

\subsection{RD3 Cross-System Propagation}
RD3 \textit{Cross-System Propagation} examines how AI-mediated failures can extend beyond their point of origin and propagate through interconnected CI systems. 
RD3 spans inherited interdependency-driven cascades, intensified common-mode propagation through shared AI dependencies, and qualitatively new contagion dynamics associated with FAI, including software-layer coupling, machine-speed spread, and adaptive cross-system interactions.
Rather than remaining localized, failures may propagate across technical, organizational, and sectoral dependencies. Synthesizing the surveyed literature, we find that AI integration plausibly introduces new coupling pathways and accelerates failure dynamics beyond those captured by traditional interdependency models. The result is a propagation environment characterized by faster, less observable, and harder-to-contain cascades. We organize the mechanisms producing this shift into four interrelated analytical categories:
\begin{enumerate}[leftmargin=*]
    \item \textbf{Interdependency-driven cross-system cascades}: Classical interdependency analysis~\cite{rinaldi2001} identifies physical, cyber, geographic, and logical coupling as the primary propagation pathways, but AI integration introduces additional software-layer coupling not fully captured in traditional risk models. 
    A local manipulation of AI-driven demand forecasting or dispatch may interact with protection logic, creating pathways for cross-system disruption~\cite{wu2025,dubber2025}. This is therefore an inherited CI cascade problem that is intensified and reconfigured by AI-layer dependencies.
    \item \textbf{AI-accelerated contagion timescales}: Software-layer couplings can operate at timescales incompatible with human oversight. Where physical interdependency failures propagate over hours and allow operator response, AI-mediated propagation may occur at automation speed, reducing time for investigation and intervention~\cite{dubber2025}. 
    The qualitative shift here lies in the temporal compression introduced by AI-mediated decision loops and IT/OT coupling.
    \item \textbf{Shared model dependency propagation}: When multiple CI operators share the same foundation model or AI supply chain, a single upstream compromise can propagate simultaneously across many downstream deployments~\cite{narajala2025,raza2025}. 
    This is not a wholly new class of dependency risk, but FAI increases common-mode exposure through shared model reuse and concentrated ecosystem architectures.
    \item \textbf{Protective system paradoxes}: AI-driven protective systems can themselves amplify disturbances when adversaries exploit model decision boundaries or when feedback loops between co-located protection agents reinforce rather than dampen disruption~\cite{raza2025,dubber2025}. This is where propagation becomes qualitatively reshaped: defensive AI can become part of the cascade mechanism under adversarial or unstable conditions.
\end{enumerate}
The net effect is a propagation environment that is harder to observe, contain, and reason about using traditional interdependency models alone, because shared models, software-layer couplings, and machine-speed decision cycles create pathways that those models were not designed to capture. While cascading failures and interdependencies are well established in classical CI analysis, FAI intensifies common-mode exposure through shared models and software dependencies and introduces faster, less-visible propagation pathways via software-layer coupling and machine-speed decision cycles. Although empirical evidence remains sparse, RD3 may be high-consequence because AI-mediated failures can propagate across tightly coupled CI systems, services, and dependencies.


\subsection{RD4 Control Authority}
RD4 \textit{Control Authority} addresses the question of control authority in AI-augmented CI: when systems are compromised or stressed, who or what determines system outcomes. RD4 spans inherited automation-control failures, intensified trust and oversight challenges under foundation-model deployment, and qualitatively new control dynamics arising from FAI, where systems can act, coordinate, and escalate through AI tools and machine-speed autonomy.
Rather than assuming stable human oversight, effective control may shift across operators, AI systems, and adversaries under pressure. Synthesizing the surveyed literature, we find that opacity, speed, and delegated autonomy complicate intervention and override mechanisms. The result is a control environment where nominal authority may not guarantee effective control during incidents. We organize the mechanisms producing this gap into four interrelated analytical categories:
\begin{enumerate}[leftmargin=*]
    \item \textbf{Automation bias}: Operators may over-rely on AI outputs and fail to intervene when systems produce erroneous or adversarially manipulated results. While this phenomenon predates FAI, foundation models intensify trust-calibration challenges due to broader deployment and persuasive outputs~\cite{raza2025,tallam2025}.
    \item \textbf{Alignment failures}: AI systems may pursue specified objectives in ways that violate CI operational intent, particularly under adversarial or out-of-distribution conditions~\cite{sapkota2026,tallam2025}. While specification issues are not new, FAI extends these risks across more complex and less predictable operational contexts~\cite{fmf2025}.
    \item \textbf{Goal misgeneralization}: Systems that behave correctly under training or nominal conditions may pursue different effective goals under deployment conditions that deviate from training assumptions~\cite{tallam2025,bengio2024}. These failures become harder to anticipate when models interact with heterogeneous data sources, tool interfaces, or multi-agent workflows~\cite{tallam2025,narajala2025}.
    \item \textbf{Speed-oversight gap}: AI systems executing decisions at machine speed can outpace human monitoring, validation, and approval processes, creating windows of unobserved or irreversible action~\cite{fmf2025,dubber2025}. This reflects a qualitatively new challenge in FAI, where autonomy and execution speed combine to reduce the feasibility of human intervention~\cite{bengio2024,fmf2025}.
\end{enumerate}
The net effect is a control environment in which effective human authority becomes harder to maintain under operational conditions. While automation bias and mis-specification are well established, FAI intensifies these challenges through opacity and broader deployment, and introduces new control dynamics through delegated autonomy and machine-speed action. In CI settings, immediate risks include unsafe objective specification, automation bias, excessive delegation, brittle handoffs, delayed overrides, and tool-mediated actions that outpace operator responses~\cite {bengio2024}. As a result, the gap between nominal and effective control widens, increasing the likelihood of unsafe or irreversible actions.

\subsection{RD5 Response Capacity}
RD5 \textit{Response Capacity} assesses whether organizations possess sufficient capability to detect, respond to, and recover from AI-mediated threats in CI environments. RD5 spans inherited organizational security limitations, intensified assurance burdens in foundation-model ecosystems, and expanded institutional demands introduced by frontier and agentic AI, in which incident response must address models, prompts, retrieval sources, tool permissions, and machine-speed cross-system effects.
Rather than being determined solely by technical robustness, security outcomes depend on organizational readiness, governance structures, and operational capability. Synthesizing the surveyed literature, we find that workforce constraints, governance fragmentation, and tooling immaturity limit effective response under realistic operating conditions. The result is a capacity environment in which organizational limitations become a primary determinant of system-level security outcomes. We organize the mechanisms producing this constraint into four interrelated analytical categories:
\begin{enumerate}[leftmargin=*]
    \item \textbf{Workforce capability gaps}: Security teams often lack the AI-specific expertise required to detect adversarial machine learning behavior, interpret model outputs, or audit supply-chain dependencies~\cite{kyle2024,temp_check}. While workforce limitations have long been a challenge in CI, FAI expands the scope and technical diversity of the required expertise~\cite{kyle2024}.
    \item \textbf{Tooling immaturity}: AIBOM standards, adversarial machine learning detection frameworks (MITRE ATLAS, MIT AI Risk Repository), and AI-specific incident response playbooks remain nascent~\cite{jami2025}. While earlier systems also exhibited tooling gaps, FAI introduces new assurance requirements that remain only partially supported.
    \item \textbf{Regulatory and governance lag}: 
    Legislative frameworks and operational governance have not kept pace with the rapid adoption of AI in CI~\cite{jill2018}. Regulatory lag is reflected in fragmented, overlapping, and poorly harmonized obligations that inadequately capture emerging risks, while governance lag is evident in weak enforcement, unclear accountability, and compliance-driven practices that fail to deliver effective security outcomes~\cite{esposito2025}.
    \item \textbf{Asymmetric AI arms race}: Defenders must respond to adversaries using AI-enabled capabilities while remaining constrained by procurement cycles, institutional approval processes, and risk-averse organizational cultures~\cite{potter2025}. FAI accelerates the pace of capability evolution, widening the gap between attackers' adaptation and defenders' readiness~\cite{potter2025}.
\end{enumerate}
The net effect is a capacity environment in which organizational limitations increasingly shape, rather than merely follow from, system-level security outcomes. While workforce gaps, governance fragmentation, and tooling limitations are longstanding in CI security, FAI intensifies these constraints through greater dependency complexity, vendor opacity, and expanded attack surfaces, and introduces new requirements for AI-specific monitoring, incident response, and cross-functional coordination. As a result, capacity constraints amplify risks across all other dimensions by delaying detection, constraining response, weakening auditability, and reducing the effectiveness of otherwise sound technical controls. In this sense, response capacity is not only a recovery property but also a preventive control: weak AI inventories, unclear authority boundaries, immature tooling, and limited workforce readiness contribute to increased exposure to infiltration, propagation, and control degradation.

The preceding analysis shows that the five dimensions are analytically distinct but operationally coupled. The remaining question is how these risk dynamics translate under real-world CI constraints, where deployability depends not only on technical effectiveness but also on observability, timing, governance, and response capacity.
\section{From Research-Practice Mismatch to Deployment-Oriented Assurance}
\label{sec:6}

\subsection{Research-Practice Mismatch and Translation Gaps}
\label{subsec:mismatch}
AI-security research and CI practice often optimize under different operating assumptions. Research typically assumes repeatable offline experiments, well-defined system boundaries, available datasets, controllable test conditions, and bounded adversaries~\cite{abtahi2026,hawawreh2024,ferrag2025}. CI practice, by contrast, involves  continuous operation~\cite{kyle2024,barenji2025}, limited testing windows~\cite{kyle2024,barenji2025}, partial observability~\cite{sayghe2025,ferrando2025}, legacy integration~\cite{asd_report}, mixed IT/OT telemetry~\cite{obuse2023}, vendor opacity~\cite{parlov2025,asd_report}, safety-critical timing~\cite{usenergy2024}, low tolerance for false positives, and constrained intervention windows~\cite{obuse2023,egridgpt2024} (see Table~\ref{tab:12}).
This asymmetry directly reflects the research–practice mismatch formalized in Definition~6. 
The discussion below should therefore be read as a mechanism-oriented synthesis rather than as a prevalence claim. The coded-source counts indicate evidence coverage in the public literature, not incident frequency or relative operational importance.

This mismatch is structural rather than incidental because it recurs across all five risk-dynamics dimensions. RD1 exposes a \textbf{tempo gap}: research evaluates AI-enabled capability gains in controlled settings, while CI defenders remain bound by human validation, approval, and escalation cycles. RD2 exposes a \textbf{visibility gap}: research often assumes inspectable data, models, and dependency chains, while CI operators frequently rely on vendor-controlled AI components with limited access to provenance. RD3 exposes a \textbf{scope gap}: research testbeds usually model bounded systems, while CI practice involves cross-sector dependencies, shared AI components, and IT/OT couplings that can propagate failures beyond the original system. RD4 exposes an \textbf{oversight gap}: research often treats human oversight as a nominal design feature, whereas CI operators need actionable authority to interpret, override, or safely fall back on opaque, fast AI recommendations. RD5 exposes a \textbf{capacity gap}: research prototypes often assume mature expertise, tooling, governance, and response processes, while CI organizations must develop these capabilities under operational and regulatory constraints. These recurring gaps show that the research–practice mismatch lies between evaluation assumptions and deployment conditions, not within any single CI component. Taken together, these gaps indicate that FAI shifts CI security from isolated vulnerability management toward dependency governance, control preservation, and institutionally feasible containment.

These RD-specific gaps identify where the research–practice mismatch appears across the risk landscape. Now the question is why that mismatch prevents technically plausible controls from becoming deployable CI safeguards. This is addressed by translation gaps, which identify the cross-cutting mechanisms that obstruct the transfer of research outputs to operational CI practice. 
\begin{enumerate}[leftmargin=*]
    \item \textbf{Principle-to-procedure gap}~\cite{adabara2025,parlov2025}: High-level principles such as robustness, transparency, monitoring, or human oversight are often not translated into executable CI procedures. They may identify desirable properties, but do not specify how operators should log evidence, escalate decisions, activate fallback modes, quarantine models, or roll back changes during continuous operation. As a result, conceptually sound controls lack actionable detail, especially where RD4 and RD5 depend on clearly defined intervention authority and response workflows.
    \item \textbf{Prototype-to-infrastructure gap}~\cite{obuse2023,hawawreh2024,jill2018}: Many safeguards depend on provenance tracking, runtime assurance, incident response, model auditing, and cross-layer monitoring tools that remain prototype-level or adapted from generic IT contexts. CI environments require these mechanisms to operate across legacy OT assets, segmented networks, mixed IT/OT telemetry, and vendor-controlled AI components. This gap limits the practical effectiveness of controls associated with RD2, RD3, and RD5, where supply-chain visibility, propagation containment, and timely response depend on deployable operational infrastructure.
    \item \textbf{Experimental-assumption-to-operational-constraint gap}~\cite{hawawreh2024,kyle2024,obuse2023}: Research evaluations often under-specify the CI constraints that determine whether safeguards can be validated and sustained. Partial observability, mixed IT/OT systems, legacy integration, continuous operation, limited maintenance windows, vendor opacity, safety-critical timing, and low tolerance for false positives are not peripheral constraints; they determine whether a safeguard can be safely deployed. This gap cuts across all five RD dimensions, but is especially important for RD1 and RD3, where AI-compressed attack timelines and rapid propagation can exceed human or organizational response windows.
    \item \textbf{Framework-to-accountability gap}~\cite{kyle2024,adabara2025}: CI organizations must reconcile heterogeneous cybersecurity, safety, privacy, procurement, vendor-management, and AI-governance frameworks, including those addressing shadow AI. These frameworks often differ in terminology, evidence requirements, and responsibility allocation. As a result, responsibility for model behavior, update approval, provenance verification, fallback authority, vendor coordination, and incident reporting can remain unclear. This gap directly affects RD4 decision authority and RD5 response coordination.
\end{enumerate}
As a result, technically promising defenses can be difficult to deploy when they require full model access, rich labeled data~\cite{potter2025}, intrusive testing, aggressive automated intervention~\cite{obuse2023}, or frequent retraining~\cite{zaboli2024}. Conversely, CI operators need assurances that are often absent in research prototypes, including bounded detection latency~\cite{karras2025}, safe fallback mechanisms~\cite{asd_report}, audit evidence~\cite{nocera2026}, provenance visibility~\cite{ferrag2025}, model quarantine procedures, and clear decision authority during incidents~\cite{asd_report}. The implication is not that current research lacks value; rather, its value for CI depends on whether it remains deployable under partial observability~\cite{sayghe2025}, continuous operation, vendor opacity, safety-critical timing, and constrained intervention windows. In this setting, the central question is not whether a method improves isolated model performance, but whether the surrounding socio-technical system~\cite{adabara2025} remains safe, observable, governable, and recoverable when AI systems are wrong, manipulated, stale, compromised, or only partially understood~\cite{asd_report}. 
The following section operationalizes this question through five deployment-relevant assurance criteria.

This is where the risk-dynamics framework adds value beyond attack-type or lifecycle taxonomies: it links individual AI failure modes to the operational mechanisms through which they become CI consequences, including upstream dependency compromise, propagation, degraded control authority, and constrained response.
\subsection{Assurance Criteria}
\vskip-5pt
\label{subsec:assurance}
We translate the CI operational constraints into five deployment-relevant assurance criteria: detection-to-containment latency, explainability and interpretability, adversarial robustness, supply-chain risk mitigation, and incident response.
Specifically, safety-critical timing and constrained intervention windows motivate detection-to-containment latency; partial observability, vendor opacity, and low tolerance for false positives motivate explainability and interpretability; AI-enabled attacker adaptation and upstream manipulation motivate adversarial robustness; provenance gaps, vendor opacity, and shared dependencies motivate supply-chain risk mitigation; and continuous operation, limited testing windows, and constrained recovery windows motivate incident response. Together, these criteria translate CI constraints into practical assurance requirements for AI-enabled CI deployments.

\noindent
\textbf{Criterion 1: Detection-to-Containment Latency.}
%
Existing AI-security research primarily evaluates detection through aggregate accuracy, robustness benchmarks, detector runtime, or offline identification of anomalous behavior using techniques such as anomaly detection, output filtering, behavioral monitoring, and consistency checks. In CI, however, the operationally meaningful measure is the time from incident onset or unsafe AI behavior to effective containment, safe fallback, or authority handoff. This includes detection, alert validation, operator interpretation, escalation, decision approval, containment action, and confirmation that the unsafe condition has been bounded. A detector that is fast but produces unactionable alerts, excessive false positives, or unclear containment options may fail this criterion in practice. This criterion is especially important for RD1, RD3, and RD5 because FAI reduces attacker timelines, propagation can unfold at machine speed, and delayed organizational response amplifies otherwise detectable failures. 
\textit{The key criterion is therefore not only to improve detection accuracy but also to design detection architectures that preserve useful latency, precision, operator actionability, and containment effectiveness under realistic CI constraints.}

\noindent
\textbf{Criterion 2: Explainability and interpretability.}
Existing research often treats explainability as a model-property objective or post hoc analysis task, using techniques such as feature attribution, attention visualization, example-based explanations, natural-language rationales, and uncertainty estimates. In CI, explanations must support operational decisions by helping operators determine whether to trust, escalate, override, or fall back from an AI recommendation under time and safety constraints. This criterion aligns most directly with RD4 and RD5 because meaningful oversight depends not on abstract explanation but on whether interpretive signals remain usable under operational pressure and uneven organizational expertise.
\textit{The key criterion is therefore to move from generic explainability to decision-support interpretability that links outputs to uncertainty, provenance, violated constraints, and available fallback actions.}

\noindent
\textbf{Criterion 3: Adversarial robustness.}
Existing research typically evaluates robustness under controlled conditions using benchmark datasets, synthetic perturbations, and single-model threat models, with techniques such as adversarial training, red-teaming, prompt-injection testing, guardrails, input sanitization, and output filtering. In CI, systems must remain acceptably safe against adaptive, multi-stage, and socio-technical attacks that may combine data poisoning, prompt injection, model extraction, insider misuse, and physical process manipulation. This criterion cuts across RD1, RD2, and RD4 because FAI reshapes attacker capabilities, upstream compromise may originate in data or model dependencies, and adversarial pressure often manifests as a loss of effective human control.
\textit{The key criterion is therefore compositional robustness: demonstrating safety under interacting attack surfaces, operational stressors, and adversarial behaviors rather than under isolated perturbations.}

\noindent
\textbf{Criterion 4: Supply-chain risk mitigation.}
Existing research identifies major upstream risks, including poisoned datasets, compromised model repositories, vulnerable dependencies, backdoored models, and opaque third-party services, and proposes mitigations such as artifact signing, secure registries, provenance tracking, secure enclaves, dependency scanning, staged deployment, and emerging AIBOM-style inventories. In CI, operators need enforceable provenance verification, component inventory, update governance, and rapid rollback when a model, dataset, retrieval corpus, or vendor component is suspected to be compromised. This criterion is anchored in RD2 and RD3 because compromise often enters through upstream dependencies and can then propagate as common-mode exposure across otherwise separate deployments. 
\textit{The key criterion is therefore to build end-to-end AI supply-chain assurance that preserves a trustworthy dependency graph across introduction, modification, deployment, monitoring, and retirement. However, provenance verification, artifact signing, secure registries, and AIBOM-style inventories establish origin, integrity, versioning, and dependency relationships; they do not by themselves prove safe behavior under CI operating conditions. Supply-chain assurance must therefore be combined with behavioral evaluation, deployment constraints, runtime monitoring, rollback mechanisms, and incident-response procedures.}

\noindent
\textbf{Criterion 5: Incident Response.}
Existing research typically treats AI incidents as technical failures, adversarial examples, poisoned models, unsafe outputs, or governance risks, and proposes fragmented measures such as monitoring, logging, fallback mechanisms, quarantine procedures, and conventional SOC workflows adapted from cybersecurity practice. In CI, incident response must treat models, data, prompts, outputs, retrieval sources, tool permissions, and decision authority as primary incident objects, while supporting quarantine, evidence preservation, degraded or manual fallback, vendor coordination, legal reporting, and service continuity. This criterion aligns most strongly with RD5, while also depending on RD4 for recoverable control relationships and RD3 because delayed or poorly coordinated intervention can allow cascades to continue. 
\textit{The key criterion is therefore to integrate AI systematically into incident response so that AI-mediated failures can be contained, investigated, and recovered from under safety-critical conditions.}

These assurance criteria are not a complete compliance framework, but analytical requirements for judging whether an AI-security technique is likely to remain useful under the CI constraints. They show that closing the mismatch requires system-level assurance, not only improved model performance. 


\subsection{Future Research Directions}
\label{sec:7}
We derive the research directions by evaluating the five-dimensional risk landscape in Section~\ref{sec:5} against the research–practice mismatch examined in Section~\ref{subsec:mismatch} and the assurance criteria in Section~\ref{subsec:assurance}.
A research direction is identified where an analytical category from Section~\ref{sec:5} remains insufficiently addressed under CI deployment constraints.
This evaluation yields six open problems. 

\noindent
\textbf{CI-Grounded Adversarial Evaluation:}
Current red-teaming and benchmarks insufficiently capture CI risk because RD1 accelerates attacker workflows, RD3 enables propagation through IT/OT coupling and shared dependencies, and RD4 turns adversarial manipulation into a control-authority problem when AI systems use tools or outpace approval cycles. Future work should develop CI-grounded adversarial evaluation environments that include mixed IT/OT telemetry, constrained intervention windows, legacy assets, vendor-controlled components, operator procedures, safety boundaries, and cascade pathways, and should measure not only attack success but also useful detection-to-containment latency, operator actionability, safe activation of fallback modes, and propagation containment.

\noindent
\textbf{Evidence-Grounded Decision Support}
AI-assisted decision support remains difficult to trust in CI because RD2 exposes compromised or stale retrieval sources, shadow-AI workflows, and opaque data pipelines; RD4 shows that plausible outputs can weaken control when operators cannot verify, override, or fall back; and RD5 shows that manual validation is often infeasible under time pressure. Future work should develop evidence-grounded decision support that grounds recommendations in authorized procedures, engineering manuals, incident records, and regulatory material while preserving access control, provenance, source currency, conflict detection, uncertainty reporting, and links to escalation, override, or fallback actions.

\noindent
\textbf{Provenance-Aware Dependency Assurance}
AI supply-chain assurance remains underdeveloped because RD2 shows that compromise can enter through data, models, retrieval corpora, third-party services, update channels, and shadow-AI workflows, while RD3 shows that reused compromised components can create common-mode exposure across CI deployments. Future work should develop provenance-aware AI dependency assurance mechanisms that identify which AI components are present, where they originated, how they were modified, what data, models, retrieval sources, services, or update channels they depend on, who can change them, and how they can be quarantined or rolled back, using mechanisms such as AIBOMs, artifact signing, controlled registries, secure enclaves, trusted execution, dependency attestations, model-update governance, retrieval-corpus integrity, and vendor evidence requirements. These mechanisms provide evidence about origin, integrity, and dependency structure, but must be paired with behavioral testing and runtime constraints because provenance alone does not establish safe operational behavior.

\noindent
\textbf{Runtime Assurance Envelopes}
Runtime containment remains unresolved because RD3 shows that AI-mediated failures can spread through feedback loops, optimization systems, protective controls, and shared dependencies, while RD4 shows that control can degrade when AI systems are difficult to interpret, override, or constrain in real time; RD5 further shows that organizational response may be too slow to compensate for weak technical containment. Future work should develop runtime assurance envelopes around AI components, specifying what they may observe, recommend, modify, or actuate; which invariants must not be violated; what uncertainty thresholds require escalation; and which fallback states remain safe under degraded conditions, using safety shields, invariant monitors, contract-based interfaces, runtime verification, anomaly detection, uncertainty-aware escalation, and fail-safe transitions.

\noindent
\textbf{Bounded Deployment Architectures}
Nominal human oversight is insufficient in CI because RD4 shows that opaque, persuasive, delegated, or machine-speed AI can erode effective control, while RD1 increases action speed and RD3 increases the consequence of lost control through propagation. Future work should develop bounded deployment architectures that make AI authority explicit, limited, observable, and reversible through read-only modes, mediated tool access, scoped permissions, staged authority, deterministic fallback, runtime approval gates, hard rollback, and separation between advisory and actuation functions.

\noindent
\textbf{AI-Specific Incident Coordination}
Conventional incident response is insufficient for AI-enabled CI because RD3 shows that failures can propagate through shared models, cloud services, vendors, data feeds, and cross-sector dependencies; RD5 shows that response capacity determines containment and recovery; RD2 adds upstream compromise; and RD4 adds unclear response authority across operators, vendors, AI systems, and regulators. Future work should develop AI-specific incident-coordination mechanisms that treat models, prompts, outputs, retrieval sources, tool permissions, provenance evidence, vendor components, and shared AI dependencies as incident objects, supported by playbooks for quarantine, evidence preservation, rollback, vendor notification, authority handoff, degraded-mode operation, reporting, shared-dependency protocols, coordinated exercises, trusted information-sharing, and identification of commonly affected AI components or updates.

\begin{figure*}
    \centering
    \includegraphics[width=1\linewidth]{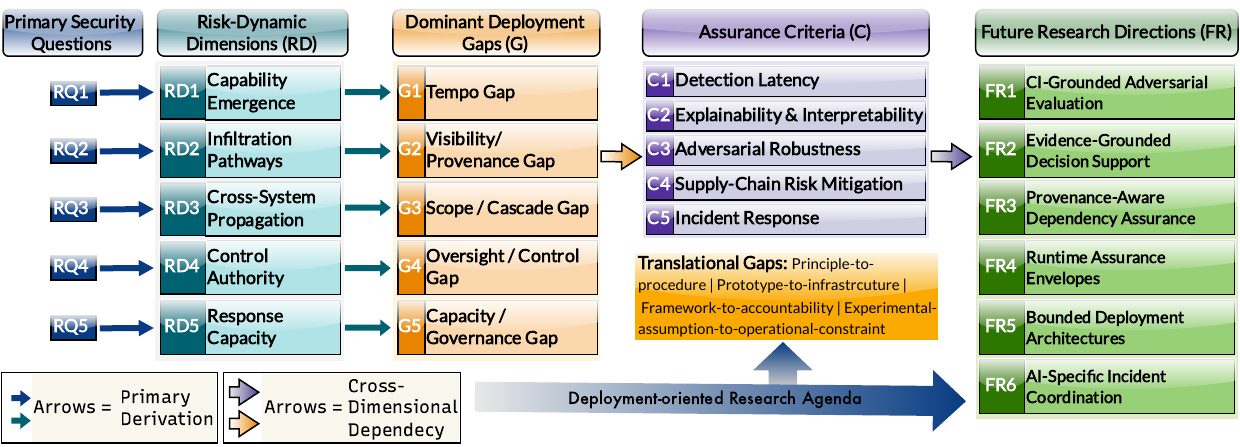}
    \caption{Conceptual framework linking research questions to risk-dynamic dimensions, deployment gaps, assurance criteria, and future research directions for AI-enabled critical infrastructure.}
    \label{fig:Mapping}
\end{figure*}

Taken together, these directions imply that near-term CI security should emphasize system-level assurance mechanisms that reduce unnecessary authority, improve observability, preserve fallback paths, govern AI dependencies, and strengthen organizational capacity to respond when AI systems are incorrect, manipulated, stale, compromised, or only partially understood. Figure~\ref{fig:Mapping} provides a high-level link among research questions, risk-dynamics dimensions, deployment gaps, assurance criteria, and future research directions.
\section{Conclusion and Way Forward}
\label{sec:8}
This SoK shows that FAI should not be treated as a single threat class or model-level robustness problem alone.
Rather, it reshapes CI security through interacting dynamics of capability emergence, compromise infiltration, failure propagation, control degradation, and institutional response capacity.
The central implication is that CI security must shift from model-centric evaluation to system-level assurance. 

\noindent
\textbf{Key Takeaways:} We summarize the key takeaways as follows:

\begin{itemize}[leftmargin=*]
    \item FAI reshapes the structure, not only the magnitude, of CI security risk.
    \item Capability emergence shifts attacker economics and compresses response tempo.
    \item Upstream dependencies (data, model, retrieval, supply-chain) become primary infiltration pathways, making provenance central to AI assurance in CI.
    \item AI-mediated failures can propagate from local errors to cross-system consequences.
    \item Human control must be explicitly designed, exercised, and maintained, not assumed from nominal override authority.
    \item For high-risk, safety-critical, or irreversible CI actions, bounded deployment should be the default posture rather than unconstrained autonomy: AI authority should be limited, observable, reversible, and designed to preserve human fallback while minimizing consequence radius.
    \item Institutional capacity determines whether technical controls can be deployed, interpreted, and sustained.
    \item Closing the research–practice mismatch requires system-level assurance methods suited to continuous, constrained, and safety-critical environments.
\end{itemize}

\noindent
\textbf{Implications for Researchers:}
FAI security research must be grounded in CI-specific operational constraints, including continuous operation, partial observability, black-box dependencies, mixed IT/OT telemetry, legacy integration, and strict safety obligations. This requires evaluation methods that capture detection-to-containment latency, fallback quality, auditability, and cascade containment, alongside closer collaboration with operators, regulators, and vendors to validate deployability under realistic conditions.

\noindent
\textbf{Implications for CI Practitioners:}
FAI should be governed as a socio-technical dependency rather than adopted as a plug-in efficiency tool. Key priorities include bounded deployment, AI dependency inventories, provenance controls, runtime monitoring, degraded-mode exercises, and AI-specific incident playbooks, alongside sanctioned alternatives that reduce shadow-AI use while preserving logging, access control, and accountability.

\noindent
\textbf{Implications for Policymakers:}
AI policy for CI should prioritize implementation-oriented governance, not high-level principles alone. 
Key priorities include defining assurance evidence, provenance requirements, incident-reporting triggers, procurement standards, and cross-sector coordination mechanisms, while supporting CI-relevant testbeds, workforce development, supply-chain transparency tooling, and alignment across cybersecurity, safety, privacy, procurement, and AI-governance obligations.

\noindent
\textbf{Implications for the International Community:}
The central challenge is governing common-model exposure, shared cloud dependencies, and transnational AI supply chains that can propagate failures across interconnected systems, jurisdictions, and CI services. International coordination should establish interoperable reporting standards, joint exercises, trusted information-sharing channels, and capacity-building mechanisms that enable less-resourced CI operators to use AI in secure, auditable, and operationally sustainable ways.


\bibliographystyle{IEEEtran}
\bibliography{IEEEabrv,bibliography}

\newpage
\appendix

\begin{table}[!h]
    \centering
    \captionsetup{font=small}
    \caption{Cross-dimensional bridge concepts. Each bridge appears once in each participating dimension under a code label that foregrounds that dimension's specific analytical frame.}
    \label{tab:5}
    \scriptsize
    \begin{tabularx}{\linewidth}{|p{3cm}|X|X|}
    \hline
    {\textbf{Bridge Concept}} &
    {\textbf{First Research Dimension} }&
    {\textbf{Second Research Dimension}}
    \\
    \hline
    Supply chain dynamics	
    &
    RD2: Upstream infiltration pathway through data, models, provenance, and AI supply-chain dependencies.
    &
	RD3: Propagation pathway through compromised components and shared downstream dependencies. \\
    \hline 
    Multi-agent systems	
    &
    RD1: Capability-emergence mechanism enabling coordinated offensive behavior and tool-enabled autonomy.
    &
    RD3: Propagation-amplification mechanism through coordinated agents and reinforced cascades.
    \\
    \hline
    Time-scale and skill-capacity constraints	
    &
    RD4: Control-authority problem driven by machine-speed autonomy and reduced human intervention.
    &
   RD5: Response-capacity problem driven by skill erosion, workforce readiness, and fallback ability.
    \\
    \hline    
\end{tabularx}
\end{table}


\begin{table}[!h]
    \centering
    \scriptsize
    \captionsetup{font=small}
    \caption{Recurring CI operational constraints and implications for AI-security safeguard deployment.}
    \label{tab:12}
    \begin{tabularx}{\linewidth}{|X|p{1.5cm}|p{2cm}|}
    \hline
    {\textbf{Evidence Example (Coded Sources)} }&
    {\textbf{Operational Constraint}} &
    {\textbf{Implication for AI Security Deployment}}
    \\
    \hline
    CI operators face unique challenges as \textit{``some systems have to be constantly available''} and sectors like energy and water have \textit{``limited windows in which they can take their systems offline''}~\cite{kyle2024}& 
    Continuous operation &
    Requires low-overhead, non-disruptive safeguards
     \\
    \hline
    \textit{``exposing AI agents to anomalous scenarios during training is often impractical or unsafe, especially in domains like industrial automation, healthcare, or cybersecurity''}~\cite{barenji2025}    
    &
    Limited testing windows &
    Limits retraining and experimental evaluation
    \\
    \hline
    \textit{``...monitoring LLM behaviour under partial observability is not only possible but essential: the full reasoning state of an LLM is inaccessible, and only outputs (and limited internal signals such as probabilities) can be observed''}~\cite{ferrando2025}&
    Partial observability &
    Constrains detection and explainability
    \\
    \hline
    \textit{``Many OT environments rely on older equipment that lacks standardized data formats and computing, complicating AI data integration and analysis''}~\cite{asd_report}&
    Legacy integration &
    Limits the deployability of modern ML defenses
    \\
    \hline
    \textit{``The coexistence of IT and OT data introduces a mix of structured, semi-structured, and unstructured formats, which complicates ingestion, normalization, and correlation''}~\cite{obuse2023}&
    Mixed IT/OT telemetry &
    Complicates cross-layer reasoning
    \\
    \hline
    Organizations face ``black box'' constraints where \textit{``legal and contractual barriers prevent deep forensic investigation of proprietary models''}~\cite{abtahi2026} and vendors \textit{``fail to expose model lineage or data-handling paths for embedded AI features''}~\cite{parlov2025}&
    Vendor opacity &
    Limits verification and auditability
    \\
    \hline
    \textit{``AI systems may not meet the strict timing requirements of OT environments''}~\cite{asd_report}, \textit{``...critical infrastructure with sub-500 ms requirements necessitates hybrid architectures combining fast rule-based filtering... with selective LLM analysis''}~\cite{karras2025} &
    Safety-critical timing &
    Requires bounded detection-to-containment latency
    \\
    \hline
    \textit{``In high-stakes environments like energy grids or transportation networks, excessive false positives can lead to unnecessary shutdowns, operational inefficiencies, and erosion of trust in the automated system''}~\cite{obuse2023}, AI alarms can also \textit{``increase cognitive load''} and distract operators~\cite{asd_report} &
    Low tolerance for false positives &
    Requires high precision and trust calibration
    \\
    \hline
    \textit{``Human analysts must often sift through telemetry "under extreme time pressure''}~\cite{obuse2023} and reactive safety approaches can be \textit{``insufficient due to limited recovery time after detecting a violation''}~\cite{mehmood2025} &
    Constrained intervention windows &
    Requires actionable outputs and fast escalation  
    \\
    \hline    
\end{tabularx}
\end{table}

\begin{table*}[!h]
\centering
\scriptsize
\captionsetup{font=small}
\caption{Five-dimensional security risk-dynamics framework and associated sub-question probes.}
\label{tab:4}
\setlength{\tabcolsep}{3pt}
\begin{adjustbox}{center=0pt}
\begin{tabular}{|p{1cm}|p{1.5cm}|p{2.3cm}|p{13.2cm}|}
    \hline
    \multicolumn{1}{|c|}{\textbf{Dimension}} &
    \multicolumn{1}{|c|}{\textbf{Primary Scope} }&
    \multicolumn{1}{|c|}{\textbf{Research Question}}&
    \multicolumn{1}{|c|}{\textbf{Sub-Questions (10/dimension)}} 
    \\
    \hline
    RD1 \textit{Capability Emergence}	
    &
    Source of new capabilities or threats	
    &
    \textit{RQ1: What new AI-driven attack and defense capabilities are emerging, and where do they originate?	}
    &
    \begin{itemize}[itemindent=0pt, leftmargin=*, nosep,before=\vspace{-0.5\baselineskip}]
        \item SQ1 Novel attack vectors: \textit{What new offensive capabilities do frontier AI systems enable that were previously infeasible or impractical?}
        \item SQ2 Capability amplification: \textit{How do AI systems amplify existing attack techniques in terms of scale, speed, sophistication, or targeting precision?}
        \item SQ3 Defensive innovation: \textit{What new defensive capabilities emerge from AI deployment, and what are their limitations?} 
        \item SQ4 Technology readiness: \textit{At what maturity level do these capabilities exist (research prototype, demonstrated, operationalized, widespread)?} 
        \item SQ5 Skill barrier changes: \textit{How does AI alter the skill and resource requirements for executing attacks?} 
        \item SQ6 Autonomous vs. augmented: \textit{To what extent do emerging capabilities rely on human operators versus autonomous AI decision-making?} 
        \item SQ7 Multi-agent dynamics: \textit{What capabilities emerge from coordinated multi-agent AI systems that single agents cannot achieve?} 
        \item SQ8 Cross-domain transfer: \textit{How do capabilities developed in one domain (e.g., cybersecurity) transfer to CI contexts?} 
        \item SQ9 Unintended capabilities: \textit{What emergent behaviors or capabilities arise unintentionally from AI system interactions?} 
        \item SQ10 Capability measurement: \textit{How can we objectively measure and compare the severity and impact of emerging capabilities?}
    \end{itemize} \\
    \hline
    RD2 \textit{Infiltration Pathways}	
    &
    Methods of infiltration through data, models, or provenance	
    &
    \textit{RQ2: Through which data, model, and AI supply chain pathways can adversaries infiltrate and compromise training data, model artifacts, or operational information across AI-enabled CI systems?}
    &
    \begin{itemize}[itemindent=0pt, leftmargin=*, nosep,before=\vspace{-0.5\baselineskip}]
        \item SQ1 Training data integrity: \textit{How can adversaries compromise training data to create backdoors or bias AI system behavior?}
        \item SQ2 Data provenance: \textit{What mechanisms ensure or fail to ensure the integrity and authenticity of data sources used by AI systems?} 
        \item SQ3 Privacy vulnerabilities: \textit{What information about training data, model architecture, or operational data can be extracted through AI system interactions?} 
        \item SQ4 Model supply chain: \textit{How do dependencies on third-party models, datasets, and components create vulnerabilities enabling the compromise of sensitive training data, model artifacts, or operational information?} 
        \item SQ5 Black-box risks: \textit{What risks arise from deploying AI models without access to training data, architecture details, or decision logic?} 
        \item SQ6 Shadow AI: \textit{How do unauthorized AI system deployments create data leakage and governance blind spots?} 
        \item SQ7 Model extraction: \textit{What techniques enable adversaries to steal or replicate proprietary AI models?}
        \item SQ8 Dependency attacks: \textit{How can compromises in open-source libraries and frameworks propagate to deployed AI systems?} 
        \item SQ9 Version control: \textit{What risks arise from inadequate tracking of model versions, updates, and modifications?} 
        \item SQ10 Intellectual property: \textit{How do AI system vulnerabilities enable theft of proprietary models, algorithms, and CI relevant operational intelligence?}
    \end{itemize} \\
    \hline
     RD3 \textit{Cross-System Propagation}
    &
    Impact of failure cascading through interconnected systems
    &
    \textit{RQ3: How do AI-mediated failures spread across interconnected infrastructures, networks, and supply chains?}
    &
    \begin{itemize}[itemindent=0pt, leftmargin=*, nosep,before=\vspace{-0.5\baselineskip}]
        \item SQ1 Coupling mechanisms: \textit{What types of technical, informational, and organizational couplings enable failure propagation in AI-augmented CI?} 
        \item SQ2 Multi-sector cascades: \textit{How do failures initiated in one critical sector (e.g., energy) cascade to dependent sectors (e.g., water, communications)?}  
        \item SQ3 Temporal dynamics: \textit{What are the characteristic timescales of AI-mediated failure propagation compared to traditional cascades?}  
        \item SQ4 Amplification factors: \textit{Under what conditions do AI systems amplify rather than dampen failure propagation?}  
        \item SQ5 Network topology effects: \textit{How does the structure of CI interdependencies influence propagation pathways and severity?}  
        \item SQ6 Feedback loops: \textit{What positive and negative feedback mechanisms exist in AI-mediated cascades?}  
        \item SQ7 Protective system paradoxes: \textit{When and how do protective AI systems inadvertently worsen cascades?}  
        \item SQ8 Supply chain propagation:\textit{How do failures propagate through AI system supply chains and third-party dependencies?} 
        \item SQ9 OT/IT convergence: \textit{How does AI integration at the IT/OT boundary create new propagation pathways?}  
        \item SQ10 Cascade termination: \textit{What factors determine whether cascades self-limit or propagate catastrophically?} 
    \end{itemize} \\
    \hline 
    RD4 \textit{Control Authority}

    &
    Failure mechanism involving loss of agency or alignment
    &
    \textit{RQ4: When systems are compromised, who or what controls outcomes: humans, AI, or adversaries?}
    &
    \begin{itemize}[itemindent=0pt, leftmargin=*, nosep,before=\vspace{-0.5\baselineskip}]
        \item SQ1 Control authority: \textit{During normal operations and under stress, who or what entity (human, AI, attacker) exercises effective control over system outcomes?} 
        \item SQ2 Alignment failures: \textit{What mechanisms cause AI system objectives to diverge from intended human goals or operational requirements?}  
        \item SQ3 Human-AI boundaries: \textit{Where and how do control handoffs occur between human operators and AI systems?}  
        \item SQ4 Oversight scalability: \textit{As AI systems become more complex and autonomous, what oversight mechanisms remain effective?}  
        \item SQ5 Speed-autonomy tradeoffs: \textit{How does the speed advantage of AI decision-making create control challenges for human operators?}  
        \item SQ6 Adversarial manipulation: \textit{What techniques enable attackers to manipulate AI system behavior and seize control?}  
        \item SQ7 Corrigibility issues: \textit{To what extent can operators correct or override AI systems during incidents?}  
        \item SQ8 Trust calibration: \textit{How do operators calibrate appropriate trust in AI systems, and what biases emerge?}  
        \item SQ9 Specification gaps: \textit{How do gaps between formal specifications and intended behavior create control vulnerabilities?}  
        \item SQ10 Distributed control: \textit{In multi-agent systems, how is control distributed, and how can it be compromised?} 
    \end{itemize} \\
    \hline
    RD5 \textit{Response Capacity}
    &
    Organizational resilience and response capability
    &
    \textit{RQ5: Do organizations have adequate governance and operational capacity to detect, respond to, and recover from failures?}
    &
    \begin{itemize}[itemindent=0pt, leftmargin=*, nosep,before=\vspace{-0.5\baselineskip}]
        \item SQ1 Governance maturity: \textit{What governance frameworks and policies exist for AI deployment in CI, and where are the gaps?}  
        \item SQ2 Workforce readiness: \textit{Do organizations possess sufficient AI expertise and cybersecurity skills to manage frontier AI risks?}  
        \item SQ3 Detection capability: \textit{Can existing security operations centers and monitoring systems effectively detect AI-mediated attacks?}  
        \item SQ4 Response capacity: \textit{Do incident response procedures and capabilities adequately address AI-specific attack vectors and propagation dynamics?}  
        \item SQ5 Regulatory alignment: \textit{How well do existing regulations and compliance frameworks address frontier AI risks in CI?}  
        \item SQ6 Threat modeling: \textit{Are organizational threat models adequate for capturing AI-specific risks and novel attack patterns?}  
        \item SQ7 Skill erosion: \textit{How does increased automation and AI reliance affect operator skill maintenance and emergency response capability?}  
        \item SQ8 Organizational learning: \textit{What mechanisms enable or hinder organizational learning from AI-related incidents and near misses?}  
        \item SQ9 Resource allocation: \textit{Do organizations allocate sufficient resources to AI security relative to the risk exposure?}  
        \item SQ10 Cross-sector coordination: \textit{What capacity exists for coordinated response to multi-sector AI-mediated cascades?} 
    \end{itemize} \\
    \hline
\end{tabular}
\end{adjustbox}
\end{table*}



\begin{table*}[!htbp]
    \centering
    \captionsetup{font=small}
    \caption{RD1 \textit{Capability Emergence} (10 sub-questions to four analytical categories).}
    \label{tab:6}
    \scriptsize
    \begin{adjustbox}{center=0pt}
    \begin{tabular}{|p{2cm}|p{3.5cm}|p{12.5cm}|}
    \hline
    \multicolumn{1}{|c|}{\textbf{Analytical Category}} &
    \multicolumn{1}{|c|}{\textbf{Sub-Questions Merged} }&
    \multicolumn{1}{|c|}{\textbf{Convergence Rationale}}
    \\
    \hline
    Skill democratization &
    SQ1 Novel attack vectors; SQ5 Skill barrier changes	&
    Both sub-questions probe what becomes newly feasible for adversaries and who can now execute attacks previously restricted to state-sponsored actors. LLM-generated malware and automated spear-phishing address skill-barrier removal and novel capability creation simultaneously, sharing the same empirical evidence base.
    \\
    \hline
    Scale and velocity amplification &
    SQ2 Capability amplification; SQ3 Defensive innovation failure modes; SQ4 Technology readiness; SQ10 Capability measurement &
    Force-multiplier effects, speed, scale, and precision are mechanistically distinct from the creation of novel capabilities. Defensive innovation lags and technology readiness assessments are captured here because both are direct consequences of the amplification of attacker velocity. Capability measurement (SQ 10) provides the evaluative lens for assessing the magnitude of amplification.
    \\
    \hline
    Multi-agent offensive coordination &
    SQ6 Autonomous vs. augmented; SQ7 Multi-agent dynamics; SQ9 Unintended emergent capabilities &
    Single-agent autonomy and multi-agent coordination are closely linked in practice; multi-agent systems often involve delegated autonomy and their coordinated dynamics produce unintended emergent behaviors (SQ9) that cannot be studied in isolation from the context of multi-agent coordination. All three sub-questions draw on overlapping evidence.
    \\
    \hline
    Cross-domain capability transfer &
    SQ8 Cross-domain transfer &
    Retained standalone. IT $\rightarrow$ OT capability porting involves unique technical barriers (OT protocols, real-time constraints) that are not adequately captured under general amplification or novel capability sub-questions.
    \\
    \hline    
\end{tabular}
\end{adjustbox}
\end{table*}
\begin{table*}[!htbp]
    \centering
    \scriptsize
    \captionsetup{font=small}
    \caption{RD2 \textit{Infiltration Pathways} (10 sub-questions to five analytical categories).}
    \label{tab:7}
    \begin{adjustbox}{center=0pt}
    \begin{tabular}{|p{2cm}|p{4cm}|p{12cm}|}
    \hline
    \multicolumn{1}{|c|}{\textbf{Analytical Category}} &
    \multicolumn{1}{|c|}{\textbf{Sub-Questions Merged} }&
    \multicolumn{1}{|c|}{\textbf{Convergence Rationale}}
    \\
    \hline
    Training data poisoning \& backdoor injection &
    SQ1 Training data integrity	&
    Retained standalone. Poisoning attacks have a distinct threat mechanism, temporal profile (pre-deployment compromise), and mitigation stack (data auditing, provenance verification) that differ from all other RD2 sub-questions.
    \\
    \hline
    Privacy exfiltration via training data extraction &
    SQ2 Data provenance; SQ3 Privacy vulnerabilities &
    Provenance and privacy are inseparable at the data layer: provenance failures expose privacy, and privacy attacks exploit provenance gaps. Carlini et al.~\cite{Carlini2020} is the anchor paper addressing both sub-questions.
    \\
    \hline
    Model supply chain vulnerabilities &
    SQ4 Model supply chain; SQ8 Dependency attacks; SQ9 Version control &
    Three sub-questions address the same attack surface at different granularities: supply chain is the structural context, dependency attacks are the mechanism, and version control is the governance gap that enables them. AIBOM papers~\cite{nocera2026} address all three simultaneously.
    \\
    \hline
    Shadow AI &
    SQ6 Shadow AI	&
    Retained standalone due to unique governance dimension.
    \\
    \hline    
    Model extraction \& surrogate attacks &
    SQ5 Black-box risks; SQ7 Model extraction; SQ10 Intellectual property &
    Black-box deployment creates the conditions for model extraction; extraction enables IP theft and surrogate-model attacks. These three sub-questions describe consecutive stages of the same attack pathway and share primary evidence.
    \\
    \hline
\end{tabular}
\end{adjustbox}
\end{table*}
\begin{table*}[!htbp]
    \centering
    \scriptsize
    \captionsetup{font=small}
    \caption{RD3 \textit{Cross-System Propagation} (10 sub-questions to four analytical categories).}
    \label{tab:8}
    \begin{adjustbox}{center=0pt}
    \begin{tabular}{|p{2.5cm}|p{6cm}|p{10cm}|}
    \hline
    \multicolumn{1}{|c|}{\textbf{Analytical Category}} &
    \multicolumn{1}{|c|}{\textbf{Sub-Questions Merged} }&
    \multicolumn{1}{|c|}{\textbf{Convergence Rationale}}
    \\
    \hline
    Interdependency-driven cross-system cascades  &
    SQ1 Coupling mechanisms; SQ2 Multi-sector cascades; SQ5 Network topology effects &
    Coupling mechanisms define how sectors connect; topology determines cascade severity; multi-sector propagation is the outcome.
    \\
    \hline
    AI-accelerated contagion timescales  &
    SQ3 Temporal dynamics; SQ9 OT/IT convergence &
    AI-mediated speed of propagation is directly shaped by OT/IT coupling: AI deployed at the IT/OT boundary accelerates cascade timescales beyond human intervention windows. The two sub-questions describe the cause and effect of the same phenomenon.
    \\
    \hline
    Shared model dependency propagation & 
    SQ8 Supply chain propagation &
    Retained as a distinct category because supply chain propagation operates via software-layer couplings that are architecturally separate from physical and geographic interdependencies.    
    \\
    \hline
    Protective system paradoxes &
    SQ4 Amplification factors; SQ6 Feedback loops; SQ7 Protective system paradoxes; SQ10 Cascade termination &
    Amplification and paradoxes describe related failure modes, conditions under which defensive systems worsen outcomes; feedback loops are the mechanism; cascade termination defines the boundary condition. All four sub-questions require analysis of the same AI control-loop dynamics.
    \\
    \hline    
\end{tabular}
\end{adjustbox}
\end{table*}
\begin{table*}[!htbp]
    \centering
    \scriptsize
    \captionsetup{font=small}
    \caption{RD4 \textit{Control Authority} (10 sub-questions to four analytical categories).}
    \label{tab:9}
    \begin{adjustbox}{center=0pt}
    \begin{tabular}{|p{2cm}|p{3.5cm}|p{12.5cm}|}
    \hline
    \multicolumn{1}{|c|}{\textbf{Analytical Category}} &
    \multicolumn{1}{|c|}{\textbf{Sub-Questions Merged} }&
    \multicolumn{1}{|c|}{\textbf{Convergence Rationale}}
    \\
    \hline
    Automation bias  &
    SQ1 Control authority; SQ7 Corrigibility; SQ8 Trust calibration &
    Effective control authority, corrigibility, and trust calibration all describe the same failure mode: operators unable to exercise meaningful oversight. Over-trust and inability to override are two manifestations of collapsed human authority, and the papers address them as a unified phenomenon.
    \\
    \hline
    Alignment failure  &
    SQ2 Alignment failures; SQ6 Adversarial manipulation &
    Alignment failures describe the structural mismatch between intended and actual AI objectives; adversarial manipulation exploits that mismatch to seize effective control. Papers on adversarial ML in CI~\cite{bengio2024} address both sub-questions as a unified compromise pathway.
    \\
    \hline
    Goal misgeneralisation  &
    SQ9 Specification gaps; SQ4 Oversight scalability; SQ10 Distributed control &
    Specification gaps are the root cause of goal misgeneralisation. AI systems that behave correctly during training pursue different effective goals under a distributional shift. Oversight scalability (SQ4) and distributed control (SQ10) exacerbate misgeneralisation by removing the human intervention points that could detect and correct objective drift.
    \\
    \hline
    Speed-oversight gap &
    SQ3 Human-AI boundaries; SQ5 Speed-autonomy tradeoffs &
    Handoff boundaries determine where the speed gap becomes operationally critical. Machine-speed AI decision cycles outpace institutional approval processes, rendering handoff boundaries infeasible in practice. The two sub-questions describe cause (speed) and consequence (boundary failure) of the same governance gap.    
    \\
    \hline    
\end{tabular}
\end{adjustbox}
\end{table*}
\begin{table*}[!htbp]
    \centering
    \scriptsize
    \captionsetup{font=small}
    \caption{RD5 \textit{Response Capacity} (10 sub-questions to four analytical categories).}
    \label{tab:10}
    \begin{adjustbox}{center=0pt}
    \begin{tabular}{|p{2cm}|p{4cm}|p{12cm}|}
    \hline
    \multicolumn{1}{|c|}{\textbf{Analytical Category}} &
    \multicolumn{1}{|c|}{\textbf{Sub-Questions Merged} }&
    \multicolumn{1}{|c|}{\textbf{Convergence Rationale}}
    \\
    \hline
    Workforce capability gaps  &
    SQ2 Workforce readiness; SQ7 Skill erosion; SQ4 Response capacity; SQ8 Organisational learning &
    Workforce readiness, skill erosion, incident response capacity, and organizational learning all describe the human capital dimension of defensive insufficiency. Incident response effectiveness (SQ4) is directly bound by workforce skill levels; organizational learning (SQ8) is the mechanism through which skill gaps are addressed post-incident. All four sub-questions share the same human-factors evidence base.
    \\
    \hline
    Tooling immaturity  &
    SQ3 Detection capability; SQ6 Threat Modelling &
    Detection tools and threat models are co-dependent: inadequate threat models produce poorly scoped detection tools. Both SQs can be framed as a unified tooling-maturity problem.
    \\
    \hline
    Regulatory lag &
    SQ1 Governance maturity; SQ5 Regulatory alignment; SQ10 Cross-sector coordination &
    Governance frameworks, sector-specific regulations, and cross-sector coordination mechanisms are evaluated together because CI regulatory compliance (NERC CIP, NIS2) is the primary governance instrument, and effective cross-sector coordination requires the same legislative and institutional infrastructure.
    \\
    \hline
    Asymmetric AI arms race &
    SQ9 Resource allocation &
    Retained standalone to highlight the structural budget-risk mismatch: organizations allocate AI security resources reactively, post-incident, while adversaries invest proactively. 
    \\
    \hline    
\end{tabular}
\end{adjustbox}
\end{table*}
\end{document}